\documentclass[pre,showpacs,10pt,amsmath,amssymb]{revtex4}
\usepackage{graphicx}

\def\'{'}

\begin{document}

\title{Multiplicative L\'evy processes: 
It\^o versus Stratonovich interpretation}

\author
{Tomasz Srokowski}

\affiliation{
 Institute of Nuclear Physics, Polish Academy of Sciences, PL -- 31-342
Krak\'ow,
Poland }

\date{\today}

\begin{abstract}
Langevin equation with a multiplicative stochastic force is considered. 
That force is uncorrelated, it has the L\'evy distribution and the power-law 
intensity. The Fokker-Planck equations, 
which correspond both to the It\^o and Stratonovich 
interpretation, are presented. They are solved for 
the case without drift and for the harmonic oscillator potential. The variance is 
evaluated; it is always infinite for the It\^o case whereas for the Stratonovich 
one it can be finite and rise with time slower that linearly, which indicates 
subdiffusion. Analytical results are compared with numerical simulations. 

\end{abstract}


\maketitle

\section{Introduction}
  
The Langevin formalism was introduced to describe motion of a particle which 
was subjected both to a Newtonian deterministic force and to the irregular 
influence of a bath of small molecules (the Brownian motion). That random force 
is uncorrelated and usually it is taken in the Gaussian form. 
Importance and wide applicability of the Gaussian distribution in 
the statistical physics follows from its stability: it constitutes 
an attractor in the functional space. According to the central limit 
theorem, a superposition of distributions with a finite variance 
leads to the Gaussian and its second moment is proportional to the time. 
However, many phenomena cannot be described in this way. 
Variance may depend on time stronger or weaker than linearly; as a consequence, 
the diffusion is anomalous \cite{bou}. Moreover, one can frequently encounter 
in nature systems far from thermal equilibrium, for which moments, in particular 
the variance, are divergent \cite{shl1,wes1,edw,bro1,tes}. 
From that point of view, the L\'evy process, which is stable but may be 
characterised by divergent moments, is an important generalisation of 
the Gaussian process. 

Divergent moments of the stochastic driving force can be attributed to nonhomogeneous 
structure of the environment, e.g. fractal or multifractal, which produces 
long-range correlations. L\'evy statistics can emerge from the temporal nature of 
the underlying process due to a subordination to the ordinary Brownian motion, 
which is highly inhomogeneous \cite{sok2}. 
The medium nonhomogeneity can enter the Langevin equation via the deterministic force. 
It is even possible to model random effects in this way (the quenched disorder 
\cite{bou,fog}). However, 
there are also processes for which the fluctuations of the stochastic force 
directly depend on the state of the system. The autocatalytic chemical reaction, 
in which the production of a molecule is enhanced by the presence of the 
molecules of the same type, can serve as an example \cite{schen}. As a result, 
the fluctuating term in the Langevin equation is multiplied by some function of the 
macroscopic variables (the multiplicative noise). Some physical problems require 
taking into account both additive and multiplicative noise \cite{den,wu}. 

In this paper, we consider the Langevin equation with the multiplicative noise, 
\begin{equation}
\label{la}
\dot x=F(x)+G(x)\eta(t), 
\end{equation}
in which $F(x)$ is the deterministic force and the uncorrelated stochastic force $\eta(t)$ 
possesses the L\'evy distribution. 
The case $G(x)=$const has been extensively studied. It has been demonstrated \cite{jes} 
that both the force-free system and that driven by the linear force are described by 
the L\'evy distribution and then the variance is divergent. The strongly non-linear 
force, however, is able to confine L\'evy flights and make the variance finite 
\cite{che,che1}. Also problems with more complicated forms of $F(x)$ were considered 
in context of Eq.(\ref{la}) with the L\'evy noise. They involve the barrier penetration 
\cite{dit} and escape from the potential well \cite{dyb2}, as well as the transport in 
a L\'evy ratchet \cite{dyb}. The general case, with multiplicative noise, can be 
regarded as a result of the adiabatic elimination of fast variables for nonlinear 
processes with additive fluctuations. Eq.(\ref{la}) can also be directly applied 
to model specific phenomena with fluctuations which are characterised by 
long tails in the distribution and a variable, in particular power-law, intensity. 
It is the case, for example, in the field of finance, where Eq.(\ref{la}) 
could be a natural generalisation of the Black-Scholes equation \cite{car1,man1}. 

In the general case of variable $G(x)$ we encounter the well-known problem 
of interpretation of the noise term in Eq.(\ref{la}) and then of 
the Stieltjes integral $\int_0^t G[x(\tau)]d\eta(\tau)$ which, in this form, 
is meaningless \cite{gar,zee}. We must decide whether 
$G[x(\tau)]$ is calculated before the noise acts, or after that. The former case 
corresponds to the It\^o interpretation \cite{ito} 
\begin{equation}
\label{ito}
\int_0^t G[x(\tau)]d\eta(\tau)=\sum_{i=1}^n G[x(t_{i-1})][\eta(t_i)-\eta(t_{i-1})], 
\end{equation}
where the interval $(0,t)$ has been divided in $n$ subintervals ($n\to\infty$). 
The above integral does not obey standard rules of the calculus, in particular 
the chain rule. Those rules are satisfied by the stochastic calculus introduced 
by Stratonovich. The stochastic integral \cite{strat} 
in this interpretation includes the process value both at the beginning and at 
the end of each subinterval: 
\begin{equation}
\label{strat}
\int_0^t G[x(\tau)]d\eta(\tau)=
\sum_{i=1}^n G\left[\frac{x(t_{i-1})+x(t_i)}{2}\right][\eta(t_i)-\eta(t_{i-1})]. 
\end{equation}
From the mathematical and technical point of view, the It\'o interpretation 
is easier to apply e.g. in the perturbation theory \cite{gar}. 
It is a suitable choice if the noise consists of clearly separated pulses, 
e.g. for a continuous description of integer processes. If, on the other hand, 
a system has some finite correlations and 
the white noise is only an approximation, the Stratonovich interpretation is more 
appropriate. It is the case if the noise has an external source, i.e. 
the noise source is not influenced by the system itself and it is possible, 
in principle, to turn off the noise \cite{vkam}. 
Possibility of applying standard rules of the calculus is an important 
technical advantage of the Stratonovich interpretation. 
It makes possible to solve the Langevin equation exactly for some nonlinear models 
with the multiplicative Gaussian white noise \cite{gra}. 
Physical predictions which follow from the Langevin equation with the noise in 
both interpretations can be qualitatively different. It is the case, 
for example, for the Ginzburg-Landau model with external
multiplicative fluctuations, which describes 
noise-induced phase transitions caused by short-term instabilities
of the disordered phase. The system exhibits that transition if one interprets 
the noise in the Stratonovich sense, but not if one uses
the It\^o interpretation \cite{car}. We demonstrate in this paper that 
predictions of both formalisms are different also for the L\'evy diffusion process. 

The aim of this paper is to study the Langevin equation (\ref{la}) with 
the multiplicative noise, which is given by the L\'evy distribution 
and the algebraic $G(x)$.  The Fokker-Planck equations (FPE) for both 
interpretations are presented in Sec.II. 
In Sec.III, the equation without external potential is solved and the case of the 
harmonic oscillator potential is discussed in Sec.IV. Results are compared with 
numerical simulations in Sec.V.

\section{Fokker-Planck equations}

In this section, we derive equations for the probability density distribution, 
which correspond to Eq.(\ref{la}) in both interpretations. The noise $\eta(t)$ 
is the symmetric L\'evy stable distribution, defined by the following expression 
\begin{equation}
\label{lev}
p_\eta(x)=\sqrt{2/\pi} \int_0^\infty \exp(-K^\mu k^\mu)\cos(kx)dk
\end{equation}
in terms of the order parameter $\mu$ ($0<\mu\le2$) and the generalised 
diffusion constant $K>0$. If $\mu<2$, $p_\eta$ has long tails: 
$p_\eta\sim |x|^{-\mu-1}$ for $|x|\to \infty$. 
The general L\'evy processes can be defined by the L\'evy-Khintchine 
formula which expresses the characteristic function in terms of the L\'evy 
measure $\nu(x)$ \cite{pro}. In the symmetric and non-Gaussian case it reads 
\begin{equation}
\label{levkhin}
\ln{\widetilde p_\eta}(k)=-t\left[\int_{|x|\ge1}(1-{\mbox e}^{ikx})\nu(x)dx+
\int_{|x|<1}(1-{\mbox e}^{ikx}+ikx)\nu(x)dx\right] 
\end{equation}
and $\nu(x)=|x|^{-\mu-1}$ corresponds to the stable process. 

In the It\^o interpretation, the probability density distribution is given by 
the fractional Fokker-Planck equation with variable diffusion coefficient 
\cite{sche,dit}
\begin{equation}
\label{fpi}
\frac{\partial}{\partial t}p_I(x,t)=-\frac{\partial}{\partial x}
F(x)p_I(x,t)+K^\mu\frac{\partial^\mu}{\partial |x|^\mu}[|G(x)|^\mu p_I(x,t)],  
\end{equation}
where the Riesz-Weyl fractional operator \cite{old} is defined in terms of the Fourier 
transform: $\frac{\partial^\mu}{\partial|x|^\mu}={\cal F}^{-1}(-|k|^\mu)$. 
Equations of the form (\ref{fpi}) can describe also jumping processes. For example, 
the fractional equation with the variable diffusion coefficient follows from 
the master equation which models the thermal activation of particles within 
the folded, heterogeneous polymers \cite{bro}; variability of the diffusion coefficient 
results there from the intrinsic potential of the monomer. Moreover, Markovian 
versions of the continuous time random walk (CTRW) produce equations 
of the form (\ref{fpi}) in the diffusion limit. 
In particular, the coupled CTRW model with a variable jumping rate, which describes 
L\'evy flights in nonhomogeneous media, involves the fractional FPE \cite{sro} 
in the form (\ref{fpi}). The diffusion term in Eq.(\ref{fpi}) 
is, in this case, the jumping rate. The drift term may also appear if we allow for 
a non-vanishing mean of the L\'evy distribution. 
The master equation describes also systems which are characterised 
by the internal noise. Those fluctuations emerge in systems of discrete particles 
and they are an inherent part of the very mechanism 
by which the state of the system evolves \cite{vkam}. A precise form of the 
deterministic equation does not exist since it is impossible for 
systems with that noise to eliminate the fluctuations. Consequently, the master 
equation describes the evolution of the entire system as a stochastic
process. 

The Stratonovich interpretation of the stochastic integral, Eq.(\ref{strat}), 
means that rules of the calculus -- the chain rule and the ordinary 
variable transformation formula -- can be applied. 
The stochastic variable can be determined by a stochastic equation with the 
additive noise which results from that with the multiplicative noise, 
obtained by a variable transformation. The above property of the stochastic 
integral can be proved on the assumption that the noise has 
a finite variance \cite{gar,zee}. If, in addition, trajectories are continuous 
(Lindeberg's condition), rules of the ordinary calculus apply to  
the Fisk-Stratonovich integral and one obtains a relation between integrals 
defined by Eq.(\ref{ito}) and (\ref{strat}): they differ only by a simple 
additive term \cite{pro}. As regards the L\'evy stable processes, they 
can be approximated by processes 
with the finite variance by introducing a cut-off in the L\'evy measure $\nu(x)$ 
in Eq.(\ref{levkhin}) (truncated L\'evy flights). Such an approximation is very 
accurate, also for a large value of the argument \cite{man}. 
In Sec.V, we will present numerical 
examples which confirm applicability of variable transformation rules of 
the ordinary calculus for the L\'evy stable processes. 

Knowing the variable transformation rules, it is possible to transform Eq.(\ref{la}) to 
a new Langevin equation, which contains the additive noise, 
instead of the multiplicative one \cite{ris}. 
For that purpose we introduce a new variable $y$ and reduce Eq.(\ref{la}) to the form: 
\begin{equation}
\label{las}
\dot y=\hat F(y)+\eta(t), 
\end{equation}
where the transformation reads
\begin{equation}
\label{tran}
y(x)=\int_0^x\frac{dx'}{KG(x')},~~~~~~~~\hat F(y)=F(x(y))\frac{dy}{dx}.
\end{equation}
The corresponding FPE is of the form 
\begin{equation}
\label{fps}
\frac{\partial}{\partial t}p_S(y,t)=-\frac{\partial}{\partial y}
\hat F(y)p_S(y,t)+\frac{\partial^\mu}{\partial |y|^\mu}p_S(y,t). 
\end{equation}
After solving the above equation, the solution of the original equation 
(\ref{la}) follows from the probability conservation rule: 
\begin{equation}
\label{tranp}
p_S(x,t)=p_S(y(x),t)\frac{dy}{dx}. 
\end{equation}

For the Gaussian case, $\mu=2$, Eq.(\ref{fps}) can be easily expressed 
in terms of the original variable $x$ and a direct relation between 
$p_S(x,t)$ and $p_I(x,t)$ can be established. 
The difference between Fokker-Planck equations for both interpretations 
resolves itself to the additional drift, $K^2G(x)G'(x)$, called 
`spurious' or `noise-induced' drift. 

In the following, we solve Eq.(\ref{la}) for two cases: without external potential 
and with the linear drift, on the assumption of both interpretations of 
the stochastic equation. We assume the diffusion coefficient in the algebraic form: 
\begin{equation}
\label{godx}
G(x)=|x|^{-\theta/\mu}. 
\end{equation}
Results can be generalised to other forms of $G(x)$ by applying the method from 
Ref. \cite{sro3}. 

\section{Force-free case}

We begin with the case of the It\^o interpretation. 
Eq.(\ref{fpi}) with $F(x)=0$ becomes the fractional diffusion 
equation with the variable diffusion coefficient, 
\begin{equation}
\label{frace}
\frac{\partial p_I(x,t)}{\partial t}=
K^\mu\frac{\partial^\mu[|x|^{-\theta} p_I(x,t)]}{\partial|x|^\mu}.
\end{equation}
The above equation results not only from the Langevin equation with the 
multiplicative noise. It constitutes the small wave number limit 
(the diffusion or fluid limit) of the master equation 
for a jumping process in the framework of 
the coupled CTRW \cite{sro}. That process is defined in terms of two probability 
distributions: of the jumping size, in the L\'evy form, and the Poissonian, 
position-dependent  waiting time distribution. 
The diffusion coefficient $|G(x)|^\mu$ in Eq.(\ref{fpi}) is then  
the jumping rate and the parameter $\theta$ governs the transport speed. 
In particular, for $\mu=2$, when the variance exists, Eq.(\ref{frace}) describes 
the anomalous diffusion process: subdiffusion for $\theta>0$, enhanced diffusion 
for $\theta<0$, and the normal diffusion for $\theta=0$. The same classification 
holds also for $\mu<2$ when we introduce fractional moments \cite{sro}. 

Eq.(\ref{frace}) can be solved in the diffusion limit by applying 
the Fox functions formalism \cite{fox,mat,sri}. 
Details of the derivation are presented in 
Refs.\cite{sro,sro1}. The Fox functions are well suited for  
problems which involve L\'evy processes since any L\'evy distribution, 
both symmetric and asymmetric, can be expressed as the function 
$H_{2,2}^{1,1}(x)$ \cite{sch}. Moreover, due to the 
multiplication rule, the term $|x|^{-\theta} p_I(x,t)$ 
in Eq.(\ref{frace}) can be easily evaluated 
and it produces the Fox function of the same order, only the coefficients are 
shifted. Those properties of Eq.(\ref{frace}) 
suggest the method of solution: we 
assume the solution in the scaling form which involve the Fox function,  
\begin{eqnarray} 
\label{s1}
p_I(x,t)=Na(t) H_{2,2}^{1,1}\left[a(t) |x|\left|\begin{array}{c}
(a_1,A_1),(a_2,A_2)\\
\\
(b_1,B_1),(b_2,B_2)
\end{array}\right.\right],
\end{eqnarray}
where $N$ is the normalization constant, 
and try to adjust the coefficients, as well as to derive the function $a(t)$. 
To implement the approximation of small $k$, we pass to the Fourier space, in which 
Eq.(\ref{frace}) reads 
\begin{equation}
\label{fracek}
\frac{\partial}{\partial t}{\widetilde p_I}(k,t)=
-K^\mu |k|^\mu{\cal F}_c[|x|^{-\theta}p_I(x,t)].
\end{equation}
According to the well-known formula, the Fourier transform from the Fox function 
is also the Fox function but of higher order: 
${\cal F}_c[H_{2,2}^{1,1}(x)]=H_{3,4}^{2,1}(k)$. 
Then we expand the Fox functions, which correspond to ${\widetilde p_I}(k,t)$ and ${\cal F}_c[|x|^{-\theta}p_I(x,t)]$, 
in the fractional powers of $k$; terms of the order $|k|^{2\mu+\theta}$ 
and higher are neglected. Both sides of Eq.(\ref{fracek}) can be adjusted 
only if all terms except $k^0$ and $|k|^\mu$ vanish. We can eliminate adverse 
terms by a proper choice of the Fox function coefficients. 
Inserting the coefficient values, determined in this way, to Eq.(\ref{frace}), 
yields the solution in the form 
\begin{eqnarray} 
\label{soli}
p_I(x,t)={\cal N}a(t)H_{2,2}^{1,1}\left[a(t)|x|\left|\begin{array}{l}
(1-\frac{1-\theta}{\mu+\theta},\frac{1}{\mu+\theta}),(a_2,A_2)\\
\\
(b_1,B_1),(1-\frac{1-\theta}{2+\theta},\frac{1}{2+\theta})
\end{array}\right.\right]; 
  \end{eqnarray}
the initial condition is $p_I(x,0)=\delta(x)$. 
Moreover we assume $\mu+\theta>0$. The coefficients $(a_2,A_2)$ 
and $(b_1,B_1)$ are essentially arbitrary and one needs additional requirements 
to settle them. Expansion of the Fox function in Eq.(\ref{soli}) yields 
$p_I(x,t)\sim |x|^{b_1/B_1}$ $(x\to 0)$. It explains why the coefficients 
$(b_1,B_1)$ have not been determined: since small $|k|$ corresponds 
to large $|x|$, the diffusion approximation does not cover the region of small $|x|$. 
In the following, we assume $b_1=\theta$ and $B_1=1$, which choice is exact 
for CTRW \cite{sro1}. The coefficients $(a_2,A_2)$, in turn, correspond 
to the shape of the distribution in the limit $\mu\to2$ \cite{sro1}. 
From Eq.(\ref{fracek}) follows a simple differential equation for $a(t)$ 
which yields $a(t)\sim t^{-1/(\mu+\theta)}$. 
The asymptotic expansion of Eq.(\ref{soli}) yields 
$p_I(x,t)\sim t^{\mu/(\mu+\theta)}|x|^{-1-\mu}$. Therefore, we obtain the L\'evy 
process which has the same distribution as the driving process, Eq.(\ref{lev}). 
The variance and all higher moments are divergent. 

For the Stratonovich interpretation, we introduce the new variable $y$, 
according to Eq.(\ref{tran}), 
\begin{equation}
\label{yodx}
y(x)=\frac{\mu}{K(\mu+\theta)}|x|^{(\mu+\theta)/\mu}\hbox{sgn}x, 
\end{equation}
which ends in Langevin equation with the additive noise. Then Eq.(\ref{fps}) reads 
\begin{equation}
\label{fps0}
\frac{\partial}{\partial t}p_S(y,t)=\frac{\partial^\mu}{\partial |y|^\mu}p_S(y,t). 
\end{equation}
It can be solved exactly and the solution expressed in the form \cite{wes,met}
\begin{eqnarray} 
\label{solsy0}
p_S(y,t)=\frac{1}{\mu|y|}H_{2,2}^{1,1}\left[\frac{|y|}{t^{1/\mu}}
\left|\begin{array}{l}
(1,1/\mu),(1,1/2)\\
\\
(1,1),(1,1/2)
\end{array}\right.\right],
\end{eqnarray}
which corresponds to the symmetric L\'evy process $y(t)$. The inverse transformation 
to the original variable yields
\begin{eqnarray} 
\label{solsx0}
p_S(x,t)=\frac{\mu+\theta}{\mu^2|x|}H_{2,2}^{1,1}\left[\frac{|x|^{1+\theta/\mu}}
{(1+\theta/\mu)(K^\mu t)^{1/\mu}}
\left|\begin{array}{l}
(1,1/\mu),(1,1/2)\\
\\
(1,1),(1,1/2)
\end{array}\right.\right]. 
\end{eqnarray}
The asymptotic expansion of the Fox function in Eq.(\ref{solsx0}) 
is given by the expression $p_S(x,t)\sim t^{\mu/(\mu+\theta)}
|x|^{-1-\mu-\theta}$ $(|x|\to\infty)$. It differs considerably 
from the It\^o result: the shape of the tail depends not only on the 
order parameter of the driving process, $\mu$, but also the $\theta$-dependence emerges. 
As a result, the variance may not be divergent. We can express the variance 
by Mellin transform of the Fox function, $\chi(s)$, and evaluate it by simple algebra: 
\begin{equation}
\label{var}
\begin{split}
\langle x^2\rangle=&2\int_0^\infty x^2 p(x,t)dx=
2\left[K(\frac{\theta}{\mu}+1)\right]^{2\mu/(\mu+\theta)}\int_0^\infty y^{2\mu/(\mu+\theta)}p(y,t)dy=\\
&\frac{2}{\mu}\left[K(\frac{\theta}{\mu}+1)\right]^{2\mu/(\mu+\theta)}t^{2/(\mu+\theta)}
\chi\left(-\frac{2\mu}{\mu+\theta}\right)=\\
&-\frac{2}{\pi\mu}\left[K(\frac{\theta}{\mu}+1)\right]^{2\mu/(\mu+\theta)}
\Gamma\left(-\frac{2}{\mu+\theta}\right)
\Gamma\left(1+\frac{2\mu}{\mu+\theta}\right)\sin\left(\frac{\pi\mu}{\mu+\theta}\right)
t^{2/(\mu+\theta)},
\end{split}
\end{equation}
where we assumed $\delta=2\mu/(\mu+\theta)<\mu$, which implies $\theta>2-\mu$. 
On that condition, the integral in 
Eq.(\ref{var}) is convergent and the variance exists. We conclude that 
the diffusion process -- in which the stochastic driving force 
is L\'evy distributed and the stochastic equation is interpreted in the Stratonovich 
sense -- may not be accelerated for the case without any external 
potential. If the variance exists, the diffusion is anomalously weak since the 
convergence condition coincides with the subdiffusion condition: the variance 
rises with time slower than linearly, $\langle x^2\rangle\sim t^\delta$. The 
slope of that dependence diminishes with the parameter $\theta$. In the case 
$\mu=2$, beside the subdiffusion, also the enhanced diffusion occurs, 
for $\theta<0$, as well as the normal one if $\theta=0$.

\section{Linear force}

In the case of stochastic motion in the harmonic oscillator field, $F(x)=-\lambda x$, 
which is governed by the Langevin equation with the Gaussian white noise 
(the Ornstein-Uhlenbeck process), the probability distribution converges with time to 
a steady state which corresponds to the Boltzmann equilibrium distribution. 
If the driving noise has the L\'evy distribution with $\mu<2$, the stationary 
limiting distribution still exist but the Boltzmann equilibrium is not reached 
and the variance is infinite \cite{jes}. We will demonstrate that, 
if the multiplicative noise is interpreted in the Stratonovich sense, 
the steady state may have the finite variance. 

In the It\^o interpretation, FPE is given by Eq.(\ref{fpi}), 
\begin{equation}
\label{fraceo}
\frac{\partial}{\partial t}p_I(x,t)=\lambda\frac{\partial}{\partial x}
[xp_I(x,t)]+K^\mu\frac{\partial^\mu}{\partial |x|^\mu}[|x|^{-\theta}p_I(x,t)], 
\end{equation}
and its Fourier transformation yields 
\begin{equation}
\label{fraceko}
\frac{\partial}{\partial t}{\widetilde p_I}(k,t)=-\lambda k\frac{\partial}{\partial k}
{\widetilde p_I}(k,t)-K^\mu |k|^\mu{\cal F}_c[|x|^{-\theta}p_I(x,t)].
\end{equation}
The solution of Eq.(\ref{fraceko}) can be obtained \cite{sro2} in a similar way as 
in the case $F(x)=0$, namely by inserting the expression (\ref{s1}) into 
Eq.(\ref{fraceko}). Then the Fourier transforms of the respective functions 
are expanded and terms of the order $|k|^{2\mu+\theta}$ and higher are neglected. 
The same conditions for the Fox function coefficients are required because 
contribution from the drift term contains only the component $|k|^\mu$. Therefore 
we obtain the solution of Eq.(\ref{fraceo}) in the form (\ref{soli}). 
The comparison of terms of the order $|k|^\mu$ on both sides of Eq.(\ref{fraceko}) 
results in a simple differential equation for the function $a(t)$. 
Its solution reads
\begin{equation}
\label{a}
a(t)=\left[\frac{\lambda/c_L}{1-\exp[-\lambda(\mu+\theta)t]}\right]^{1/(\mu+\theta)}.   
\end{equation}
The constant $c_L=K^\mu h_0/\mu h_\mu$ involves the expansion coefficients 
$h_\mu$ and $h_0$ 
of the functions ${\widetilde p_I}$ and ${\cal F}_c[|x|^{-\theta}p_I]$, which 
correspond to the orders $|k|^\mu$ and $k^0$, respectively. 
They are given by $h_\mu=N(\mu+\theta)\Gamma(-\mu)
\Gamma(1+\mu+\theta)\cos(\pi\mu/2)/\Gamma[a_2+A_2(1+\mu)]
\Gamma[-(\mu+\theta)/(2+\theta)]$ and 
$h_0=N(\mu+\theta)/(2+\theta)\Gamma[a_2+A_2(1-\theta)]$, 
where the normalization constant 
$N=\Gamma[-\theta/(2+\theta)]\Gamma[a_2+A_2]/2\Gamma(1+\theta)\Gamma[-\theta/(\mu+\theta)]$.
The numerical values of the solution can be computed 
by means of series expansions, both for small and large $|x|$. Expansion 
of the function (\ref{soli}) in powers of $|x|^{-1}$ 
produces the following expression
\begin{equation}
\label{expd}
p_I(x,t)=N(\mu+\theta)\sum_{i=1}^\infty\frac{\Gamma[1+(\mu+\theta)i]}
{\Gamma(a_2+A_2[1-\theta+(\mu+\theta)i])\Gamma(-\frac{\mu+\theta}{2+\theta}i)i!}
a(t)^{\theta-(\mu+\theta)i}|x|^{-1+\theta-(\mu+\theta)i}, 
\end{equation}
where $a(t)$ is given by Eq.(\ref{a}). In the following, we assume the remaining 
Fox function coefficients in the form: $a_2=1/2+\theta(1+\theta)/(2+\theta)$ and 
$A_2=1-1/(2+\theta)$. For these coefficients, $p_I(x,t)$ agrees with the exact diffusion 
equation solution in the case $\mu=2$: it corresponds to the stretched-Gaussian 
dependence \cite{sro1} 
\begin{equation}
\label{expo}
p_I(x,t)\sim a^{1+\theta}|x|^\theta \exp[-\hbox{const} (a|x|)^{2+\theta}]. 
\end{equation} 
Eq.(\ref{expd}) implies the asymptotic shape of the distribution, 
$p_I(x,t)\sim |x|^{-1-\mu}$, the same as for the case $F(x)=0$. Therefore, 
the variance is also divergent for all $\theta$ and $\mu<2$. 

\begin{figure}[tbp]
\includegraphics[width=8.5cm]{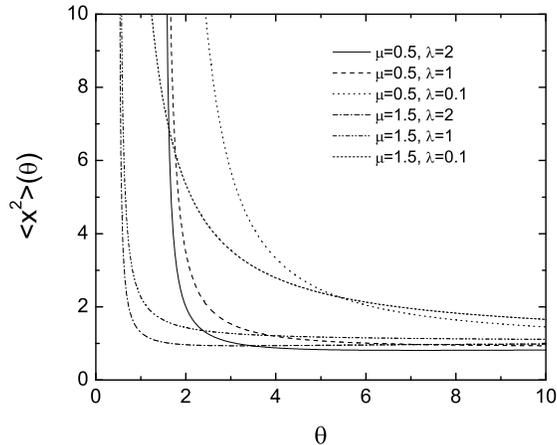}
\caption{Variance of the steady-state solution of Eq.(\ref{la}) 
in the Stratonovich interpretation for the case $F(x)=-\lambda x$, calculated 
from Eq.(\ref{var1}).}
\end{figure}

\begin{figure}[tbp]
\includegraphics[width=8.5cm]{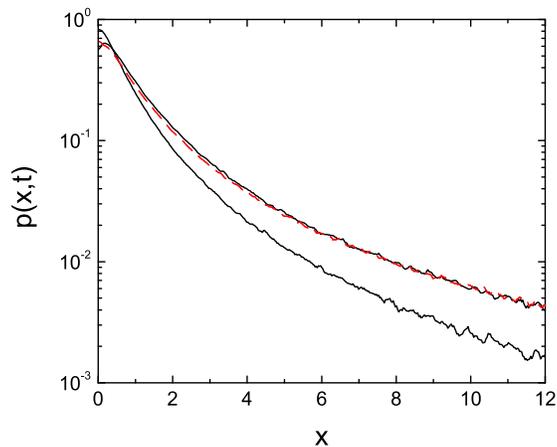}
\caption{(Colour online) Comparison of numerical solutions of Eq.(\ref{la}) 
for the case $G(x)=x$ and $F(x)=0$, with $\mu=1.5$ at $t=1$. 
The distribution $p_I(x,t)$ (lower line) 
was calculated according to Eq.(\ref{sti}) and $p_S(x,t)$ from Eq.(\ref{heun}) 
(upper line). The dashed line presents result of the evaluation 
of $p_S(x,t)$ by means of Eq.(\ref{strat}). 
}
\end{figure}

\begin{figure}[tbp]
\includegraphics[width=8.5cm]{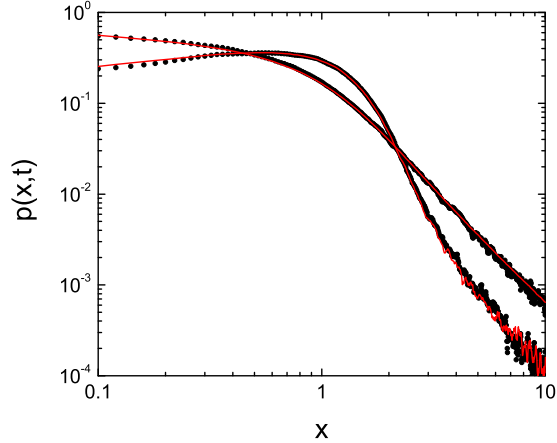}
\caption{(Colour online) Comparison of numerical solutions of Eq.(\ref{la}) 
for $G(x)=|x|^{-\theta/\mu}$ and $F(x)=-\lambda x$, with parameters: 
$\mu=1.8$, $\lambda=1$ and $t=1$, 
in the Stratonovich interpretation. Distributions marked by lines were calculated 
by using the variable transformation, from Eq.(\ref{heun}), and those marked by dots 
-- directly from Eq.(\ref{strat}). Cases for two values of $\theta$ are presented: 
$\theta=-0.2$ (upper line for small and large $x$) and $\theta=0.5$. 
}
\end{figure}

To obtain the solution of Eq.(\ref{la}) in the Stratonovich interpretation, we 
introduce the variable $y$, which is given by Eq.(\ref{yodx}), and transform the drift 
term according to Eq.(\ref{tran}). The Langevin equation takes the form 
\begin{equation}
\label{laso}
\dot y=-\frac{\lambda}{\mu}(\mu+\theta)y+\eta(t) 
\end{equation}
and FPE, expressed by the new variable, has the constant diffusion coefficient:
\begin{equation}
\label{fraces}
\frac{\partial}{\partial t}p_S(y,t)=\lambda(1+\theta/\mu)\frac{\partial}{\partial y}
[yp_S(y,t)]+\frac{\partial^\mu}{\partial |y|^\mu}p_S(y,t).  
\end{equation}
The above equation can be solved exactly with the initial condition 
$p_S(y,0)=\delta(y)$ \cite{jes}. The Fourier transform of the solution reads 
${\widetilde p}_S(k,t)=\exp(-K^\mu\sigma(t)|k|^\mu)$, where
\begin{equation}
\label{as}
\sigma(t)=\frac{1-\exp[-\lambda(\mu+\theta) t]}{\lambda(\mu+\theta)}. 
\end{equation}
After inverting the Fourier transform and transforming back to the original 
variable, we obtain the probability density distribution in the following form 
\begin{eqnarray} 
\label{solsx1}
p_S(x,t)=\frac{\mu+\theta}{\mu^2|x|}H_{2,2}^{1,1}\left[\frac{|x|^{1+\theta/\mu}}
{(1+\theta/\mu)K\sigma^{1/\mu}}
\left|\begin{array}{l}
(1,1/\mu),(1,1/2)\\
\\
(1,1),(1,1/2)
\end{array}\right.\right], 
\end{eqnarray}
which can be evaluated by series expansions, similar to Eq.(\ref{expd}). 
The tail of the distribution $p_S(x,t)$ has the same form as in the case $F(x)=0$: 
$p_S(x,t)\sim |x|^{-1-\mu-\theta}$. The second moment is convergent 
if $\theta>2-\mu$. On that condition, the system reaches with time a steady 
state which is characterised by the variance 
\begin{equation}
\label{var1}
\lim_{t\to\infty}\langle x^2\rangle(t)=
-\frac{2}{\pi\mu}\left[K(\frac{\theta}{\mu}+1)\right]^{2\mu/(\mu+\theta)}
\Gamma\left(-\frac{2}{\mu+\theta}\right)
\Gamma\left(1+\frac{2\mu}{\mu+\theta}\right)\sin\left(\frac{\pi\mu}{\mu+\theta}\right)
[(\mu+\theta)\lambda]^{-2/(\mu+\theta)}.
\end{equation}
The above quantity is presented in Fig.1 as a function of $\theta$ for some values of $\mu$ and 
$\lambda$. In all cases it declines with $\theta$; this fall is particularly rapid for large 
$\lambda$. Predominantly, the distribution is broader for smaller $\mu$ but this trend turns to 
the opposite in the limit $\theta\to\infty$. The parameter $\theta$ influences 
the convergence speed to the steady state, according to Eq.(\ref{as}), which is the same 
as for the It\^o interpretation, cf. Eq.(\ref{a}). 

\begin{figure}[tbp]
\includegraphics[width=8.5cm]{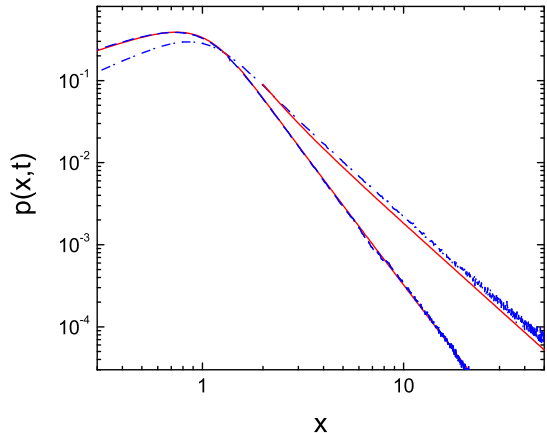}
\caption{(Colour online) Probability distributions obtained from trajectory 
simulations for the case $F(x)=-\lambda x$. The Stratonovich ($p_S(x,t)$) 
and It\^o ($p_I(x,t)$) results are marked by the dashed and dashed-dot lines, respectively. 
The analytical solutions of Eq.(\ref{la}) (solid lines) follow from the series expansions 
of Eq.(\ref{solsx1}) both for small and large $|x|$, as well as from Eq.(\ref{expd}). 
Parameters are the following: $\mu=1.2$, $\theta=1$, $\lambda=1$, and $t=1$.}
\end{figure}

\section{Numerical simulations}

In this section, we compare the analytical results with numerical stochastic trajectory 
simulations from the Langevin equation (\ref{la}). 
For the It\^o interpretation we apply the Euler method. Eq.(\ref{ito}) implies that 
for each integration step $\tau$ the stochastic integral from the function $G(x)$ can be 
expressed by the noise value $\eta_i$ by means of the following formula \cite{wer} 
\begin{equation}
\label{sti}
\int_0^t G(x(t))\eta dt=\int_0^t G(x(t))d\eta=\sum_{i=0}^N G(x(t_i))\tau^{1/\mu}\eta_i, 
\end{equation}
where $t_i=i\tau$ and $N=t/\tau$; the random numbers $\eta_i$ are sampled from the L\'evy 
distribution (\ref{lev}) with the order parameter $\mu$. 
In the case of the Stratonovich interpretation, we first transform Eq.(\ref{la}) 
to the corresponding equation with the additive noise. Then we apply 
the Heun method of integration, 
\begin{equation}
\label{heun}
y_{i+1}=y_i+[\hat F(y(t_i))+\hat F[y(t_i)+\hat F(y(t_i))\tau]]\tau/2+\tau^{1/\mu}\eta_i.
\end{equation}
Transformation back to the original variable produces the stochastic trajectory $x(t)$. 
We demonstrate results of those algorithms in Fig.2 for the simple case $G(x)=x$ and 
$F(x)=0$. The applied parameters, number of trajectories $10^6$ and $\tau=0.001$, 
ensure sufficient accuracy. The above results are compared with the probability 
distribution obtained from the numerical integration of the equation with the 
multiplicative noise, in which the stochastic integral is defined by Eq.(\ref{strat}). 
Agreement of both results is very good. A similar comparison is presented in Fig.3 for 
the nonlinear $G(x)$ in the form (\ref{godx}) with two values of the parameter 
$\theta$, both positive and negative; no cut-off in the L\'evy measure was introduced. 
This case is numerically more complicated because the difference formula is not 
explicit and numerical solving of a nonlinear equation is required at each integration 
step. Agreement of both methods of calculation in Fig.3 demonstrates that 
rules of the ordinary calculus are applicable for the L\'evy processes 
with infinite variance. 
\begin{figure}[tbp]
\includegraphics[width=8.5cm]{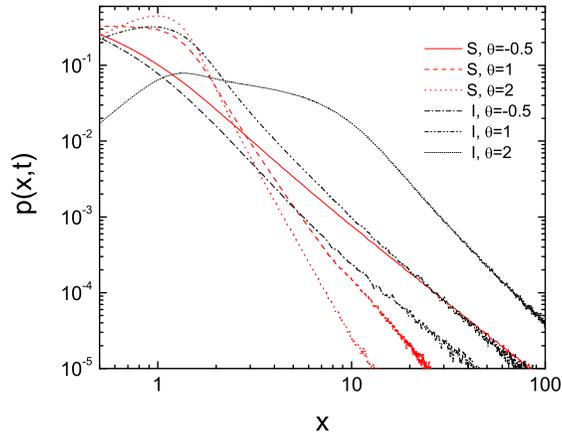}
\caption{(Colour online) Probability distributions $p_S(x,t)$ and 
$p_I(x,t)$ obtained from trajectory simulations for the case 
$F(x)=-\lambda x$ at $t=1$, with $\mu=1.5$ and $\lambda=1$. 
}
\end{figure}

The comparison with analytical results for the linear drift, presented in Fig.4, 
indicate a good agreement of those methods of solution. 
On the other hand, results which correspond to the It\^o and Stratonovich 
interpretations for the same case differ considerably. 
Since they coincide for $\theta=0$, one can expect that the difference 
rises with $|\theta|$. In Fig.5, the probability 
distributions for various $\theta$, evaluated by numerical trajectory simulations, 
are presented. The difference between $p_I(x,t)$ and $p_S(x,t)$ for $\theta=2$ is 
indeed very large. The slope of $p_I(x,t)$ 
remains constant for a given $\mu$ and that of $p_S(x,t)$ rises with $\theta$. 
We present those slopes, as a function of the order parameter $\mu$, in Fig.6. 
The slopes rise with $\mu$, according to the analytical results $-\mu-1$ 
and $-\mu-\theta-1$ for the It\^o and Stratonovich interpretations, respectively.  

\begin{figure}[tbp]
\includegraphics[width=8.5cm]{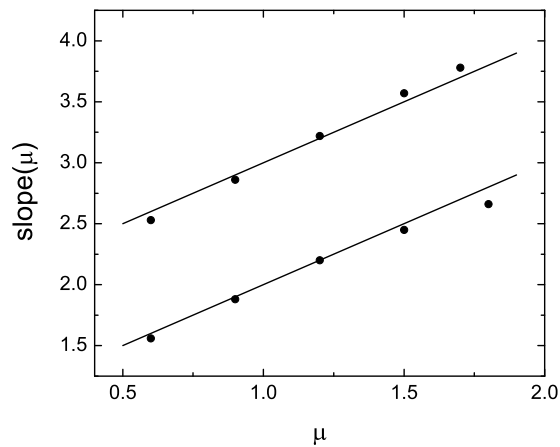}
\caption{Slope of the solution $p_I(x,t)$ (lower points) and of $p_S(x,t)$ 
(upper points) for large $|x|$, calculated by the numerical integration of Eq.(\ref{la}) 
for the case $F(x)=-\lambda x$ and $t=1$. The parameters: $\lambda=1$ and $\theta=1$. 
The lower and upper straight lines have the slopes 
$-\mu-1$ and $-\mu-\theta-1$, respectively.}
\end{figure}

\section{Summary and conclusions} 

We have studied the non-linear Langevin equation with the multiplicative white noise 
which is distributed according to the L\'evy statistics and has the power-law strength. 
The corresponding Fokker-Planck equation is fractional and its form depends on the 
interpretation of the stochastic equation. In the It\^o interpretation, 
FPE possesses the variable diffusion coefficient. In the case without any external force, 
FPE coincides with the diffusion equation which was obtained in the framework of the 
coupled CTRW with the position-dependent waiting time. Solution of FPE 
with variable diffusion coefficient, 
both for the case without drift and for the harmonic oscillator potential, 
represents the L\'evy process with simple scaling and the same order parameter as the 
driving noise. This property does not hold for other potentials. For example, 
solution to the problem of the wedge-shaped potential, $F(x)\sim\hbox{sgn}x$, studied 
in Ref.\cite{sro2}, is a combination of two L\'evy processes and 
simple scaling is lacking. Nevertheless, slowly decaying L\'evy tails are present 
in that solution and then the variance is divergent for all $\theta$. 

In the Stratonovich interpretation, FPE has been obtained 
by the variable transformation; in this case the problem is reduced to 
Langevin equation with the additive noise. The resulting probability distributions 
differ considerably from those in the It\^o sense. In particular, shape of the tail 
depends on the parameter of noise intensity $\theta$ and, 
as a result, the variance may be convergent. Therefore the diffusion process, 
for the case without drift, can be either accelerated or anomalously weak, in 
contrast to the It\^o result, which predicts only the accelerated diffusion. 
For the case $\mu=2$, both approaches differ by an additional, effective 
force in the Langevin equation. The disagreement between both interpretations 
is in fact not surprising since the deterministic force in the Langevin equation
is not just the Newtonian one; those forces are identical only in the case of the external 
noise, i.e. for the Stratonovich interpretation \cite{vkam}. Since the difference 
between the interpretations is deterministic in nature no different underlying 
stochastic dynamics is implied. For the general L\'evy stable 
processes, relation between distributions in both interpretations is more complicated 
and the corresponding equations cannot be related one to the other 
by means of a drift term. 

The distribution tails, which are algebraic and fall rapidly enough 
to ensure convergence of the variance (`fat tails'), are of physical importance. 
They are well known e.g. in the field of the economic research \cite{gab,ple}. 
Based on the poor empirical performance of the Black-Scholes model of option 
pricing, which mathematically resolves itself to the Langevin equation, one proposes
to replace the Gaussian noise by the a L\'evy one but with a truncated tail \cite{car1}. 
Such process may converge to the Gaussian so slowly, that numerical calculations 
yield only fat tails \cite{sro1,sok}. The Langevin equation with the multiplicative 
L\'evy noise in the Stratonovich interpretation could be an alternative model of the 
fat tails.


\begin{thebibliography}{99}

\bibitem{bou}
J.-P. Bouchaud and A. Georges, Phys. Rep. {\bf 195}, 127 (1990).

\bibitem{shl1}
M. F. Shlesinger, G. M. Zaslavsky, and U. Frisch, {\it L\'evy Flights 
and Related Topics in Physics}, Lecture Notes in Physics, Vol.450 
(Springer-Verlag, Berlin, 1995). 

\bibitem{wes1}
B. J. West and W. Deering, Phys. Rep. {\bf 246}, 1 (1994). 

\bibitem{edw} 
A. M. Edwards {\it et al.}, Nature (London) {\bf 449}, 1044 (2007). 

\bibitem{bro1}
D. Brockmann, L. Hufnagel, and T. Geisel, 
Nature {\bf 439}, 462 (2006).

\bibitem{tes}
C. J. Tessone, M. Cencini, and A. Torcini, 
Phys. Rev. Lett. {\bf 97}, 224101 (2006).

\bibitem{sok2}
I. M. Sokolov, Phys. Rev. E {\bf 63}, 011104 (2000).

\bibitem{fog}
H. C. Fogedby, Phys. Rev. Lett. {\bf 73}, 2517 (1994).

\bibitem{schen}
A. Schenzle and H. Brand, Phys. Rev. A {\bf 20}, 1628 (1979). 

\bibitem{den}
S. I. Denisov, A. N. Vitrenko, and W. Horsthemke, 
Phys. Rev. E {\bf 68}, 046132 (2003).

\bibitem{wu}
Wu Da-jin, Cao Li, and Ke Sheng-zhi, 
Phys. Rev. E {\bf 50}, 2496 (1994).

\bibitem{jes}
S. Jespersen, R. Metzler, and H. C. Fogedby, 
Phys. Rev. E {\bf 59}, 2736 (1999). 

\bibitem{che}
A. Chechkin, V. Gonchar, J. Klafter, R. Metzler, and L. Tanatarov, 
Chem. Phys. {\bf 284}, 233 (2002). 

\bibitem{che1}
A. V. Chechkin, V. Yu. Gonchar, J. Klafter, R. Metzler, and L. V. Tanatarov, 
J. Stat. Phys. {\bf 115}, 1505 (2004). 

\bibitem{dit}
P. D. Ditlevsen, Phys. Rev. E {\bf 60}, 172 (1999). 

\bibitem{dyb2}
B. Dybiec, E. Gudowska-Nowak, and P. H\"anggi, 
Phys. Rev. E {\bf 75}, 021109 (2007). 

\bibitem{dyb}
B. Dybiec, E. Gudowska-Nowak, and I. M. Sokolov, 
Phys. Rev. E {\bf 78}, 011117 (2008). 

\bibitem{car1}
A. Cartea and D. del-Castillo-Negrete, Physica A {\bf 374}, 749 (2007). 

\bibitem{man1}
R. N. Mantegna and H. E. Stanley, J. Stat. Phys. {\bf 89}, 469 (1997).

\bibitem{gar}
C. W. Gardiner, {\it Handbook of Stochastic Methods for Physics, Chemistry
and the Natural Sciences} (Springer-Verlag, Berlin, 1985).

\bibitem{zee}
Z. Schuss, {\it Theory and Applications of Stochastic Differential Equations} 
(John Wiley \& Sons, New York, 1980). 

\bibitem{ito}
K. It\^o, Proc. Imp. Acad. Tokyo {\bf 20}, 519 (1944). 

\bibitem{strat}
R. L. Stratonovich, SIAM J. Control {\bf 4}, 362 (1966). 

\bibitem{vkam}
N. G. van Kampen, J. Stat. Phys. {\bf 24}, 175 (1981). 

\bibitem{gra}
R. Graham and A. Schenzle, Phys. Rev. A {\bf 25}, 1731 (1982). 

\bibitem{car}
O. Carrillo, M. Iba\~nes, J. Garc\'ia-Ojalvo, J. Casademunt, and J. M. Sancho, 
Phys. Rev. E {\bf 67}, 046110 (2003).

\bibitem{pro}
P. E. Protter, {\it Stochastic Integration and Differential Equations} 
(Springer-Verlag, Berlin, 2005). 

\bibitem{sche}
D. Schertzer, M. Larchev\^{e}que, J. Duan, V. V. Yanovsky, and S. Lovejoy, 
J. Math. Phys. {\bf 42}, 200 (2001). 

\bibitem{old}
K. B. Oldham and J. Spanier, {\it The Fractional Calculus}, (Academic Press, San Diego, 1974).

\bibitem{bro}
D. Brockmann and T. Geisel, Phys. Rev. Lett. {\bf 90}, 170601 (2003).

\bibitem{sro}
T. Srokowski and A. Kami\'nska, Phys. Rev. E {\bf 74}, 021103 (2006).

\bibitem{man}
R. N. Mantegna and H. E. Stanley, Phys. Rev. Lett. {\bf 73}, 2946 (1994).

\bibitem{ris}
H. Risken, {\it The Fokker-Planck Equation} (Springer-Verlag, Berlin,
1996).

\bibitem{sro3}
T. Srokowski, Phys. Rev. E {\bf 78}, 031135 (2008).

\bibitem{fox}
C. Fox, Trans. Am. Math. Soc. {\bf 98}, 395 (1961).

\bibitem{mat}
A. M. Mathai and R. K. Saxena, {\it The $H$-function with Applications in Statistics 
and Other Disciplines} (Wiley Eastern Ltd., New Delhi, 1978).

\bibitem{sri}
H. M. Srivastava, K. C. Gupta, and S. P. Goyal, 
{\it The $H$-functions of one and two variables with applications} 
(South Asian Publishers, New Delhi, 1982). 

\bibitem{sro1}
T. Srokowski, Physica A {\bf 388}, 1057 (2009). 

\bibitem{sch}
W. R. Schneider, in {\it Stochastic Processes in Classical and Quantum Systems, 
Lecture Notes in Physics}, edited by S. Albeverio, G. Casati, D. Merlini 
(Springer, Berlin, 1986), Vol. 262.

\bibitem{wes}
B. J. West, P. Grigolini, R. Metzler, and T. F. Nonnenmacher,
Phys. Rev. E {\bf 55}, 99 (1997). 

\bibitem{met}
R. Metzler and J. Klafter, Phys. Rep. {\bf 339}, 1 (2000).

\bibitem{sro2}
T. Srokowski, Phys. Rev. E {\bf 79}, 040104(R) (2009).

\bibitem{wer}
A. Janicki and A. Weron, {\it Simulation and Chaotic Behavior of
$\alpha$-Stable Stochastic Processes} (Marcel Dekker, New York,
1994).

\bibitem{gab}
X. Gabaix, P. Gopikrishnan, V. Plerou, and H. E. Stanley,
Nature {\bf 423}, 267 (2003).

\bibitem{ple}
V. Plerou, P. Gopikrishnan, L. A. Nunes Amaral, M. Meyer, and H. E. Stanley, 
Phys. Rev. E {\bf 60}, 6519 (1999). 

\bibitem{sok}
I. M. Sokolov, A. V. Chechkin, and J. Klafter, 
Physica A {\bf 336}, 245 (2004). 

\end{thebibliography}
\end{document}